\documentclass[submission,copyright,creativecommons]{eptcs}
\providecommand{\event}{LFMTP 2024} 

\usepackage{iftex}
\usepackage{proof}
\usepackage{amsmath}
\usepackage{amssymb}
\usepackage{xcolor}
\usepackage{listings}
\usepackage{relsize}

\ifpdf
  \usepackage{underscore}
  \usepackage[T1]{fontenc}
\else
  \usepackage{breakurl}
\fi

\newcommand{\ie}{{\it i.e.}}

\newcommand{\etal}{{\it et al.}}

\newcommand{\perm}{$\sim$}
\newcommand{\munion}{{\tt ++}}

\newcommand{\lambdax}[3]{\lambda #1:#2.#3}
\newcommand{\app}{\ }
\newcommand{\ra}{\rightarrow}
\newcommand{\lam}{\backslash}
\newcommand{\arrow}{\pftext{arrow}}

\colorlet{lprolog}{blue!70!black}
\colorlet{abellatop}{blue!70!green}
\colorlet{abellatac}{orange!30!black}
\colorlet{abellabad}{red!80!yellow}

\lstdefinelanguage{lprolog}{%
  alsoletter={-},
  classoffset=0,%
  morekeywords={sig,module,type,kind,pi,sigma,end},%
  keywordstyle=\color{lprolog},%
  classoffset=0,%
  otherkeywords={:-,=>,<=,\&},%
  sensitive=true,%
  morestring=[bd]",%
  morecomment=[l]\%,%
  morecomment=[n]{/*}{*/},%
}

\lstdefinelanguage{abella}[]{lprolog}{%
  alsoletter={-},
  classoffset=1,%
  morekeywords={Close,CoDefine,Define,Kind,Query,Quit,Specification,
    Set,Split,Theorem,Type,Undo,by,as,prop,true,false,forall,exists,nabla,
    Context,with,elems},%
  keywordstyle=\color{abellatop},%
  classoffset=2,%
  morekeywords={abbrev,apply,backchain,case,coinduction,cut,
    induction,inst,intros,monotone,on,permute,rename,left,right,witness,
    search,split,to,unabbrev,unfold,assert,with,subst,distr,lift,into,over},%
  keywordstyle=\color{abellatac},%
  classoffset=3,%
  morekeywords={undo,abort,skip,clear},%
  keywordstyle=\color{abellabad}\underbar,%
  classoffset=0,%
}

\newcommand{\pftext}[1]{\mbox{\lstinline[basicstyle=\ttfamily,keepspaces=true]|#1|}}

\title{Binding Contexts as Partitionable Multisets in Abella}

\author{Terrance Gray \qquad\qquad Gopalan Nadathur
\institute{University of Minnesota, Minneapolis, MN 55455, USA}
\email{grayx501@umn.edu \quad\qquad\qquad ngopalan@umn.edu\quad\,}
}

\def\titlerunning{Binding Contexts as Partitionable Multisets in Abella}
\def\authorrunning{T. Gray \& G. Nadathur}

\begin{document}

\maketitle

\begin{abstract}
When reasoning about formal objects whose structures involve binding,
it is often necessary to analyze expressions relative to a context
that associates types, values, and other related attributes with
variables that appear free in the expressions. We refer to such
associations as binding contexts. Reasoning tasks also require
properties such as the shape and uniqueness of associations concerning
binding contexts to be made explicit. The Abella proof assistant,
which supports a higher-order treatment of syntactic constructs,
provides a simple and elegant way to describe such contexts from which
their properties can be extracted. This mechanism is based at the
outset on viewing binding contexts as ordered sequences of
associations. However, when dealing with object systems that embody
notions of linearity, it becomes necessary to treat binding contexts
more generally as partitionable multisets. We show how to adapt the
original Abella encoding to encompass such a generalization. The key
idea in this adaptation is to base the definition of a binding
context on a mapping to an underlying ordered sequence of associations. We
further show that properties that hold with the ordered sequence view
can be lifted to the generalized definition of binding contexts and
that this lifting can, in fact, be automated. These ideas find use
in the extension currently under development of the two-level logic
approach of Abella to a setting where linear logic is used as the
specification logic. 
\end{abstract}

\section{Introduction}\label{intro}

It is often necessary to develop specifications and to reason about
formal objects whose structures incorporate some notion of binding.
Examples of such objects include formulas, types, proofs, and
programs.
A recursive analysis of such objects requires the examination of their
subparts in which there may be occurrences of free variables.
This analysis is usually parameterized by an association of some kind,
such as a type, a value, or a property, with each of these variables.
This paper concerns support for such associations, which we refer to
as {\it binding contexts}, in reasoning tasks.

The focus of our work is the treatment of binding contexts relative
to a particular reasoning system, the Abella proof assistant~\cite{baelde14jfr}.
A defining characteristic of Abella is that it provides intrinsic
support for the higher-order approach to abstract syntax.
At the representation level, this support derives from the use of the
terms of the simply typed lambda calculus as the means for encoding
objects.
At the level of the logic, Abella incorporates the special generic
quantifier $\nabla$, pronounced as \emph{nabla}, to move binding into
the meta-level and the associated nominal constants to encode free
variables. 
Further, it allows properties of binding contexts to be made explicit
through the definition of \emph{context predicates} and \emph{context
relations} and thereby to be used in proofs.

While Abella provides rich support for working with binding contexts,
one aspect that it does not treat adequately with respect to these
contexts is \emph{linearity}.
This requirement arises, for instance, when bound variables take on
the connotation of resources that must be used exactly once within the overall syntactic object.
To provide support for this viewpoint, it becomes necessary to encode
binding contexts as partitionable entities.
We show in this paper how this capability can be built into the
Abella system.
The key idea underlying our proposal is to view binding contexts as
multisets that are \emph{permutation invariant} and that can be
constructed from two simpler multisets through multiset union.
Thus, if \perm\ is an infix operator representing the permutation
relation between multisets and \munion\ is an infix multiset union
operator, the expression $G$ \perm\ $(G_1$ \munion\ $G_2)$ encodes the
fact that $G_1$ and $G_2$ partition the multiset 
$G$.\footnote{Partitioning of multisets can be described without the
use of a permutation relation; this is mainly a convenient way to do it
if we have the relation.}

Unfortunately, the ability to partition a multiset is not by itself
sufficient for the usual reasoning tasks.
When $G_1$ and $G_2$ have been determined to be partitions of a
binding context $G$, we need also to know that each of them
independently satisfies the properties needed to be the required kind of
binding context.
A related issue is that we must be able to define what it means to be
a binding context in a particular reasoning task when these contexts
may be constructed using multiset unions.
A major part of our work here is to outline a systematic method for
realizing these requirements.
Our proposal in a nutshell is to identify what it means to be a
binding context through the definition of a context predicate or
relation while initially viewing it as an ordered sequence or list of
associations.
This definition can then be lifted to arbitrary multisets through the
permutation relation.
Distributivity of the property over multiset union then factors
through the same permutation relation.
An auxiliary consequence of what we show is the fact that this scheme
is to a substantial extent automatable.

The rest of the paper is structured as follows.
In the next section, we identify in more detail the idea of binding
contexts and describe their realization in Abella when they are
represented in a list-based form; we assume in this presentation, and,
indeed, the rest of the paper, a familiarity with the Abella system.
Section~\ref{sec:partition} then identifies the need for linearity with
respect to binding contexts in specifications and the additional
constructors and definitions that suffice to realize it.
Of course, it still remains to be shown how to make things work at the
reasoning level.
Section~\ref{sec:lifting} explains how context predicates can be defined
when binding contexts may be constructed using the multiset union
operator and how the properties of such contexts can be extracted into
lemmas even in this situation.
Section~\ref{sec:ctxrels} shows that these ideas extend also to the
setting of context relations, which embody the simultaneous
description of multiple correlated contexts.
Section~\ref{sec:auto} discusses a schematic presentation of context
predicates and context relations and explains how the lifting
procedure may be automated, describing some tactics
for implementing the corresponding algorithms.
Section~\ref{sec:related} discusses related work and Section~\ref{sec:concl}
closes out the paper by sketching the use of our work in the
particular application domain that has motivated it.

\section{Binding Contexts and their Conventional Treatment in Abella}\label{sec:bindingctx}

Towards understanding the nature of binding contexts and the kinds of
properties that must be associated with them in reasoning tasks, we
may consider the example of type assignment for the simply typed
lambda calculus.
We limit the expressions in the calculus to those constructed from
variables using the operations of application, written as
$(e_1\app e_2)$, and abstraction, written as $\lambdax{x}{\tau}{e}$. 
The rules for associating types with expressions in this calculus are
then the following:

\begin{center}
\begin{tabular}{ccc}

  \infer{\Gamma \vdash x : \tau}
        { x : \tau \in \Gamma}

  \quad
  &
  \quad
  
  \infer{\Gamma \vdash (e_1\app e_2) : \tau}
        {\Gamma \vdash e_1 : \tau' \ra \tau \qquad \Gamma \vdash e_2 :
          \tau'}

  \quad
  &
  \quad
  
  \infer[x\ \mbox{new to}\ \Gamma]
        {\Gamma \vdash \lambdax{x}{\tau'}{e} : \tau' \ra \tau}
        {\Gamma, x : \tau' \vdash e : \tau}
\end{tabular}
\end{center}

Type assignment for closed terms is ultimately a relation between a
term and a type.
However, a recursive definition of this relation requires us to
consider type assignments to open terms under the assumption that the
free variables in the term have designated types.
Thus, the relation must be formalized as a ternary one, written as
$\Gamma \vdash e : \tau$.
In this example, $\Gamma$ constitutes the {\it typing} or {\it binding context}.
The structure of $\Gamma$ is governed by the rule for assigning types
to abstractions.
Based on this rule, we can observe some properties that are implicitly
associated with $\Gamma$: it assigns types only to variables and there 
is at most one assignment to any variable.
While $\Gamma$ is built one element at a time and seems to have the
structure of an ordered sequence, we are free to think of it a
multiset or even a set. 
Note finally that a {\it closed-world assumption} applies
to the rules: a term may be assigned a type only by virtue of these
rules.

Relational presentations of the kind above have a natural translation
into Abella specifications.
To present this in the particular example under consideration, we must
first describe a representation for the types and terms of the
simply typed lambda calculus.
We will use the Abella types \pftext{ty} and \pftext{tm} for encodings of
expressions in these two categories.
We will also use the constant \arrow\ of type $\pftext{ty}\ra
\pftext{ty}\ra \pftext{ty}$ to represent the function type
constructor, and the constants 
\pftext{app} and \pftext{abs}, respectively of type
$\pftext{tm} \ra \pftext{tm}\ra \pftext{tm}$ and
$\pftext{ty}\ra (\pftext{tm} \ra \pftext{tm}) \ra \pftext{tm}$, to
represent application and abstraction in the object language.
Note the use of a higher-order abstract syntax
representation here; for example, the term
$\lambdax{x}{\tau_1 \ra \tau_2}{\lambdax{y}{\tau_1}{x\app y}}$ in
the object language would be encoded by the Abella term
$(\pftext{abs}\app
    (\arrow\app \overline{\tau_1}\app \overline{\tau_2}) \app
    (\pftext{x} \lam \pftext{abs} \app
                        \overline{\tau_1}\app
                        (\pftext{y} \lam \pftext{app}\app\pftext{x}
                                                      \app \pftext{y})))$,
where $\overline{\tau_1}$ and $\overline{\tau_2}$ are the
representations of the types $\tau_1$ and $\tau_2$,
respectively.\footnote{We recall that abstraction is written as the
infix operator $\backslash$ in Abella, \ie, the
expression $\lambda x.F$ is denoted by $x \backslash F$.}
In this context, the content of the type assignment rules is captured
by the following Abella declarations that ultimately provide an
inductive definition for the ternary type assignment relation
\pftext{type\_of}: 
\begin{lstlisting}
Kind ty_assoc type.
Type ty_of tm -> ty -> ty_assoc.

Define member : A -> list A -> prop by
member X (X :: L) ;
member X (Y :: L) := member X L.

Define type_of : list ty_assoc -> tm -> ty -> prop by
type_of G X T := member (ty_of X T) G ;
type_of G (app M N) T := exists T', type_of G M (arrow T' T) /\ type_of G N T';
type_of G (abs T E) (arrow T T') := nabla x, type_of (ty_of x T :: G) (E x) T'.
\end{lstlisting}

Focusing on the first argument of the \pftext{type\_of} relation, we
see that it has the kinds of properties that we observed of
binding contexts that arise in type assignment and that it represents.
While it has the structure of an Abella list, it can equally be viewed
as a multiset or a set; the use of the \pftext{member} predicate
relative to it is compatible with all these views.
In the intended use of the predicate, this collection is constructed one
item at a time via the clause for assigning types to abstractions.
The use of the \pftext{nabla} quantifier also ensures that the
associations it provides pertain only to nominal constants---which 
represent the free variables in object language terms in the
logic---and that there is at most one such association in it for any 
such constant.

The properties we have described for binding contexts in type
assignment can be important to reasoning tasks.
They are key, for example, to showing the uniqueness of type
assignment to any typeable term: the proof of this fact hinges on 
the observations that the typing context does not assign types to
applications or abstractions and that the assignments to variables are 
unique.
However, in a formalized setting, it is not enough that these
properties hold.
It must also be made explicit that they do.
This can be done in Abella by what are commonly referred to as {\it
  context definitions}.
In the example in question, the following definition serves this
purpose:
\begin{lstlisting}
Define ty_ctx : list ty_assoc -> prop by
ty_ctx nil ;
nabla x, ty_ctx (ty_of x T :: G) := ty_ctx G.
\end{lstlisting}
The \pftext{nabla} quantifier in the head of the second clause is to
be understood as follows: it must be instantiated by a nominal
constant in generating an instance of the clause and the substitutions
for \pftext{T} and \pftext{G} that generate the instance must not
contain that constant.
The formula \pftext{(ty\_ctx G)} now serves to assert that \pftext{G} is
a typing context with the necessary properties.

It is useful to drill down a little on the last statement.
One of the requirements of a typing context is that it associates
types only with nominal constants, \ie, the representatives of
variables in terms.
The following definition identifies the predicate \pftext{name} as a
recognizer for such constants:
\begin{lstlisting}
Define name : A -> prop by
nabla n, name n.
\end{lstlisting}
Using it, we can capture the desired property in the following Abella
theorem about the ``shape'' of the entities comprising a typing context:
\begin{lstlisting}
Theorem ty_ctx_mem : forall L X,
ty_ctx L -> member X L -> exists n T, name n /\ X = ty_of n T.
\end{lstlisting}
Another property that is important is the uniqueness of type
association.
This can be rendered into the following Abella theorem:
\begin{lstlisting}
Theorem ty_ctx_uniq : forall L X T1 T2,
ty_ctx L -> member (ty_of X T1) L -> member (ty_of X T2) L -> T1 = T2
\end{lstlisting}
These theorems can both be proved by induction on the definition of
\pftext{ty\_ctx}.
Once we have these properties, it is an easy matter to prove the
following theorem:
\begin{lstlisting}
Theorem ty_uniq : forall L X T1 T2,
ty_ctx L -> type_of L X T1 -> type_of L X T2 -> T1 = T2.
\end{lstlisting}
The uniqueness of type assignment for closed terms follows easily from
this theorem.

Although our discussion in this section has been oriented around an
example, the underlying concepts are quite general.
Binding contexts manifest themselves commonly in specifications about
syntactic constructs that incorporate binding notions. 
Context definitions make explicit the structure of such
contexts when a higher-order abstract syntax representation is used
for the constructs.
The properties that we must extract from such definitions to support
other reasoning tasks take two forms.
First, there are {\it membership lemmas} like \pftext{ty\_ctx\_mem}
that constrain the shape of the elements of the context.
Second, there are {\it uniqueness lemmas} like \pftext{ty\_ctx\_uniq}
that assert the uniqueness of bindings.
We have seen how context definitions can be played out and the
associated lemmas can be proved when contexts are limited to being
constructed and analyzed one item at a time.
We will next show why this view of the structure of
contexts needs to be generalized and then demonstrate how such a
generalization may be accommodated. 

\section{Partitionable Binding Contexts and Multiset Union}\label{sec:partition}

The treatment of binding contexts that we have described in the
previous section does not support the aspect of
\emph{linearity} that is relevant to some applications. 
An example of such an application is provided by the simply
typed \emph{linear} lambda calculus.
To be well-formed, terms in this calculus must have the additional
property that every bound variable is used exactly once.
Under this restriction, the term
$\lambdax{x}{\tau_1 \ra \tau_2}{\lambdax{y}{\tau_1}{x\app y}}$
is well-formed but 
$\lambdax{x}{\tau_1 \ra \tau_1 \ra \tau_2}
            {\lambdax{y}{\tau_1}{x\app y \app y}}$ and 
$\lambdax{x}{\tau_1}{\lambdax{y}{\tau_2}{y}}$
are not.

If we are to build a linearity check into the type assignment process,
the rule for assigning a type to an application must incorporate the
idea of \emph{partitioning} a binding context.
To support this possibility, we propose allowing binding contexts
to be constructed using one other operation, that of \emph{multiset
union}. 
More specifically, we shall continue to use the type \pftext{(list A)}
to represent such contexts but now we will interpret this type as that
of multisets of elements of type \pftext{A} rather than that of
ordered sequences. 
We will continue to use the constant \pftext{nil} and the infix
operator \pftext{::} as constructors of this type, but now
interpret the latter as a means for adding an element to a
multiset.
Additionally, the type now has one other constructor, the infix
operator \pftext{++} of type \pftext{(list A) -> (list A) -> (list A)}.
An expression of the form \pftext{G1++G2} is intended to
represent a multiset whose elements comprise those of
\pftext{G1} and \pftext{G2}.

The \pftext{member} predicate must be adapted to this
changed syntax.
Its definition becomes the following:
\begin{lstlisting}
Define member : A -> list A -> prop by
member X (X :: G) ;
member X (Y :: G) := member X G ; 
member X (G1 ++ G2) := member X G1 \/ member X G2.
\end{lstlisting}
To accommodate linearity, we will need a
counterpart to this predicate that represents the \emph{selection} of
a member from a multiset that simultaneously yields a smaller
multiset. 
Towards this end, we will use a predicate called \pftext{select} that
has the definition below.
\begin{lstlisting}
Define select : A -> list A -> list A -> prop by
select X (X :: G) G ;
select X (Y :: G) (Y :: G') := select X G G' ;
select X (G1 ++ G2) (G1' ++ G2) := select X G1 G1' ;
select X (G1 ++ G2) (G1 ++ G2') := select X G2 G2'.
\end{lstlisting}

We want to be able to treat binding contexts that have the same
elements as equivalent, regardless of how they are constructed.
Towards this end, we introduce a permutation predicate \pftext{perm}
for multisets. It is useful to define this, at a high-level, by
recursion on the number of elements in each context, defining an
auxiliary \pftext{no\_elems} predicate that holds of a context that is
empty.
The relevant definitions follow: 
\begin{lstlisting}
Define no_elems : list A -> prop by
no_elems nil ;
no_elems (G1 ++ G2) := no_elems G1 /\ no_elems G2.

Define perm : list A -> list A -> prop by
perm G1 G2 := no_elems G1 /\ no_elems G2 ;
perm G1 G2 := exists X G1' G2',
    select X G1 G1' /\ select X G2 G2' /\ perm G1' G2'.
\end{lstlisting}
The auxiliary definition of \pftext{no\_elems} also gives us a means
of ensuring that all bound variables are used in a specification of a
linear system.
We can assert that the context satisfies this
predicate after we have analyzed the entirety of a term to ensure  
no variables were introduced by an abstraction but left unused.

We introduce a convenient notational shorthand for the
predicate \pftext{perm}: we shall write \pftext{G1~G2} to represent
\pftext{(perm G1 G2)}.
The \pftext{perm} predicate and the \pftext{++} operator together give
us a means for encoding a partition  of $n$ multisets into $m$
multisets, which we can write as  
\pftext{G1++...++Gn~D1++...++Dm}.
Note that the permutation component of this expression allows elements
to be distributed in any order between the multisets on the other
side---so that partitioning does not depend on the elements to
partition having been ordered correctly ahead of time. 

The components that we have described in this section provide us the
necessary means for writing linear specifications.
Let us bring this out through the definition of a typing relation for
the linear lambda calculus that only assigns types to valid linear
lambda terms.
The definition of this relation, which we denote by the predicate
\pftext{ltype\_of}, is as follows: 
\begin{lstlisting}
Define ltype_of : list ty_assoc -> tm -> ty -> prop by
ltype_of G X T := exists G', select (ty_of X T) G G' /\ no_elems G' ;
ltype_of G (app M N) T := exists T' G1 G2, 
    G ~ G1 ++ G2 /\ ltype_of G1 M (arrow T' T) /\ ltype_of G2 N T' ;
ltype_of G (abs T E) (arrow T T') := 
    nabla x, ltype_of (ty_of x T :: G) (E x) T'.
\end{lstlisting}
It is worth mentioning the differences between the definition of this
predicate and that of \pftext{type\_of} in
Section~\ref{sec:bindingctx} to understand how the linearity
constraints are enforced.
The use of \pftext{select} in the first clause ensures that a
particular association for a bound variable cannot be used more than
once, and the \pftext{no\_elems} assertion ensures that every
association must have been used.
The formula \pftext{G~G1++G2} realizes a partitioning of
the context \pftext{G} and thereby ensures that the type assignment to
a particular bound variable must be used for typing exactly one of the 
two subcomponents of an application.
The structure of the last clause, which is unchanged from the
definition of \pftext{type\_of}, still ensures that the binding
context has associations only for variables and that an association for
any variable is unique.
However, we must reason now about the effect of partitioning to see
that these properties actually hold.
 
\section{Reasoning About Binding Contexts in the Generalized Form}\label{sec:lifting}

In proving properties of relations whose definitions involve binding
contexts in the extended form, we will once again need to establish
membership and uniqueness lemmas pertaining to the binding contexts.
For example, in showing the uniqueness of type assignment as expressed
by the \pftext{ltype\_of} relation, we will need the counterparts of
the \pftext{ty\_ctx\_mem} and \pftext{ty\_ctx\_uniq} lemmas for
contexts in the new form.
We show here how this can be done.
The difficulty that must be addressed is that the introduction of
multiset union breaks the view of contexts being constructed one
element at a time.
The solution that we propose is based on flattening a context with
arbitrary structure into one that is constructed in the conventional
way.
We present the idea relative to an example but its generality should
be clear from the discussion.

\subsection{Lifting Context Definitions to the Generalized Form}

In Section~\ref{sec:bindingctx}, we defined the predicate
\pftext{ty\_ctx} to make explicit the logical structure of typing
contexts for the simply typed lambda calculus.
This definition must now be extended to cover contexts that are
constructed using the multiset union operator.
We might think of doing this by adding a third clause akin to the
following to the definition of \pftext{ty\_ctx}:
\begin{lstlisting}
ty_ctx (G1 ++ G2) := ty_ctx G1 /\ ty_ctx G2.
\end{lstlisting}
Unfortunately, this idea does not work: such a clause would break the
property of the binding context that associations for a particular
name are unique, as nothing in it enforces that the names associated
within \pftext{G1} and \pftext{G2} are distinct from each other, even
if they are distinct within each  individual context.

The insight that underlies the solution that we propose is that the
properties in question should not depend on the order in which the
associations in a binding context are listed or the way in which they
are distributed over a multiset, only on what those associations are.
Thus, it would suffice if we could restructure the multiset
construction and rearrange its elements so as to produce a form that
satisfies the \pftext{ty\_ctx} predicate that we had defined earlier.
Further, the kind of projection that is necessary here can be
accomplished through the \pftext{perm} predicate that
relates two multisets with possibly different structures so long as
they have the same elements.
Thus, in the present example, the context definition might be given by
the \pftext{ty\_ctx'} predicate that is defined as follows:
\begin{lstlisting}
Define ty_ctx' : list ty_assoc -> prop by
ty_ctx' G := exists L, G ~ L /\ ty_ctx L.
\end{lstlisting}
This predicate applies to typing contexts presented in the generalized
form since the \pftext{perm} predicate is defined over multisets that
could include the \pftext{++} operator as well.
Note, however, the definition of \pftext{ty\_ctx} is dependent on an
``ordered sequence'' view, \ie, the given context must be projected
onto one in this form to assess whether it possesses the necessary
properties. 

\subsection{Proving Membership and Uniqueness Lemmas}

The new definition must still enable us to prove lemmas about the
shape of the associations in the binding context as well as their
uniqueness.
These lemmas are the following in the present situation:
\begin{lstlisting}
Theorem ty_ctx_mem' : forall G X,
ty_ctx' G -> member X G -> exists n T, name n /\ X = ty_of n T.

Theorem ty_ctx_uniq' : forall G X T1 T2,
ty_ctx' G -> member (ty_of X T1) G -> member (ty_of X T2) G -> T1 = T2.
\end{lstlisting}

The proofs of these lemmas also embody a process of ``lifting'' of
properties established based on the ordered sequence view through the
projection.
First observe that the theorems \pftext{ty\_ctx\_mem} and
\pftext{ty\_ctx\_uniq} continue to hold despite the change in the
definition of the \pftext{member} predicate.
Specifically, the definition of this predicate reduces to the original
one when the multiset argument is limited to having a list-like
structure, a structure that is forced by the \pftext{ty\_ctx}
predicate. 
But now we can also prove the following (generic) lemma that
states that membership in a multiset is preserved through a permutation:
\begin{lstlisting}
Theorem mem_replace : forall X G G', member X G -> G ~ G' -> member X G'. 
\end{lstlisting}
Since contexts described by \pftext{ty\_ctx'} are only a permutation away from
those described by \pftext{ty\_ctx}, this is sufficient to lift the theorems
\pftext{ty\_ctx\_mem} and \pftext{ty\_ctx\_uniq} into \pftext{ty\_ctx\_mem'} and
\pftext{ty\_ctx\_uniq'}.
We need only apply the lemma to replace the \pftext{member}
predicates in one theorem with those in the other.

\subsection{Distributivity of Context Properties over Multiset
  Unions}

The multiset union constructor was introduced originally to facilitate
a partitioning of contexts.
For this to be useful for the intended purpose, the facet of being a
context of the desired kind must distribute over such partitioning.
In our example, this translates into the desire that
the following theorem be provable:
\begin{lstlisting}
Theorem ty_ctx_distr : forall G G1 G2,
ty_ctx' G -> G ~ G1 ++ G2 -> ty_ctx' G1 /\ ty_ctx' G2.
\end{lstlisting}

Once again, we can prove the desired property by establishing a corresponding
property for list-like contexts, and then lifting that property to contexts that may
include the multiset union operator in their formation. One approach to stating the
first property involves defining a predicate that encodes an ordered partition 
relation between three lists:
\begin{lstlisting}
Define partition : list A -> list A -> list A -> prop by
partition nil nil nil ;
partition (X :: L) (X :: L1) L2 := partition L L1 L2 ;
partition (X :: L) L1 (X :: L2) := partition L L1 L2.
\end{lstlisting}
By exploiting the fact that the relative order of elements in a list
\pftext{L} is preserved within the related lists \pftext{L1} and
\pftext{L2}, we can easily prove the following theorem that states
that the property of being a typing context is preserved by such
partitions: 
\begin{lstlisting}
Theorem ty_ctx_distr_part : forall L L1 L2,
ty_ctx L -> partition L L1 L2 -> ty_ctx L1 /\ ty_ctx L2.
\end{lstlisting}

We can lift this theorem to \pftext{ty\_ctx'} and \pftext{perm}-style
partitions by relating \pftext{partition} and \pftext{perm}.
Towards this end, we first define a predicate that captures the
property that a context has a list-like structure:
\begin{lstlisting}
Define is_list : list A -> prop by
is_list nil ;
is_list (X :: L) := is_list L.
\end{lstlisting}
The following lemma then provides the necessary bridge:
\begin{lstlisting}
Theorem perm_to_part : forall L G1 G2,
is_list L -> L ~ G1 ++ G2 -> exists L1 L2, 
    G1 ~ L1 /\ G2 ~ L2 /\ partition L L1 L2.
\end{lstlisting}
Essentially, the lemma says that a partition of the elements in a list
into two arbitrary contexts can be flattened into a \pftext{partition} 
between lists of the same elements.
It can be proved by inverting the permutation and using the elements
extracted from the multisets \pftext{G1} and \pftext{G2} to construct
the lists \pftext{L1} and \pftext{L2}.
The proof relies critically on the following lemma which allows
elements in a multiset \pftext{G} that is related by \pftext{perm} to
\pftext{L} to be extracted one at a time in the order they appear 
in \pftext{L}:
\begin{lstlisting}
Theorem sel_replace : forall X G1 G1' G2,
G1 ~ G2 -> select X G1 G1' -> exists G2', G1' ~ G2' /\ select X G2 G2'. 
\end{lstlisting}
Note that this lemma is, in fact, a counterpart to \pftext{mem\_replace} for
\pftext{select}.

At this stage, we have all the ingredients in place to prove the
\pftext{ty\_ctx\_distr}  theorem.
Given any context \pftext{G} for which \pftext{ty\_ctx'} holds, there must, 
by definition, be an \pftext{L} such that \pftext{G~L} and
\pftext{ty\_ctx L}.
Since \pftext{G~G1++G2} holds, by properties of \pftext{perm}, it 
must then be the case that \pftext{L~G1++G2} holds.
Now, using theorems \pftext{perm\_to\_part} and
\pftext{ty\_ctx\_distr\_part}, we can conclude  that there are
contexts \pftext{L1} and \pftext{L2} such that \pftext{G1~L1},  
\pftext{G2~L2}, \pftext{ty\_ctx L1}, and \pftext{ty\_ctx L2}
hold; we will need to  show that \pftext{is\_list L} holds in order to
invoke theorem \pftext{perm\_to\_part}, but this follows easily from
the fact that \pftext{ty\_ctx L} holds.
Using the definition of \pftext{ty\_ctx'}, it is then immediate that
\pftext{ty\_ctx' G1} and  \pftext{ty\_ctx' G2} must hold.

\section{Generalization to Context Relations}\label{sec:ctxrels}

Typical meta-theoretic reasoning tasks require us to relate different
kinds of analyses over the same object-language expression.
When the expression embodies binding constructs, these analyses would
be parameterized by binding contexts.
In the Abella setting, the shape of each of these contexts must be 
characterized by a definition.
When different analyses are involved in the property to be proved,
there will generally be an additional requirement: the
content of the different binding contexts parameterizing the analyses
must be coordinated in an appropriate way.
\emph{Context relations} constitute the canonical mechanism in Abella
for phrasing context definitions to suit the reasoning needs in such
situations.
The generalized multiset structure is needed for dealing with
linearity in this situation as well and the methods for supporting it
bear a remarkable resemblance to those when only one binding context
is involved.
We bring this observation out in this section through an example.

The example we consider is that of relating typing judgments across
a translation.
The target language for the translation shall be the linear variant of the
simply typed lambda calculus that we introduced in
Section~\ref{sec:partition}.
The source language, which we will call mini linear ML, shall be  
similar, except that it shall include an additional \pftext{let}
construct.
To represent such expressions, we introduce the constant \pftext{let}
that has the type
$\pftext{ty}\ra \pftext{tm} \ra (\pftext{tm} \ra \pftext{tm}) \ra
\pftext{tm}$. 
Observe that higher-order abstract syntax is used again in the
encoding of \pftext{let} expressions: the expression
$\pftext{let X:}\tau\ \pftext{=}\ \pftext{V}\ \pftext{in}\ \pftext{F}$ is represented by
$(\pftext{let}\app\overline{\tau}\app
\overline{\pftext{V}}\ \pftext{(X}\backslash
\overline{\pftext{F}}\pftext{)})$, where $\overline{\tau}$,
$\overline{\pftext{V}}$, and $\overline{\pftext{F}}$ are the
representations of $\tau$, \pftext{V}, and \pftext{F}, respectively.  
The typing relation for the source language is now given by the
following definition:
\begin{lstlisting}
Define mltype_of : list ty_assoc -> tm -> ty -> prop by
mltype_of G X T := exists G', select (ty_of X T) G G' /\ no_elems G' ;
mltype_of G (app M N) T := exists T' G1 G2, 
    G ~ G1 ++ G2 /\ mltype_of G1 M (arrow T' T) /\ mltype_of G2 N T' ;
mltype_of G (let T' V E) T := exists G1 G2, 
    G ~ G1 ++ G2 /\ mltype_of G1 V T' /\ 
    nabla x, mltype_of (ty_of x T' :: G2) (E x) T ; 
mltype_of G (abs T E) (arrow T T') := 
    nabla x, mltype_of (ty_of x T :: G) (E x) T'.
\end{lstlisting}

\smallskip
The translation of mini linear ML expressions to the linear lambda
calculus essentially replaces \pftext{let} expressions by applications.
It is formalized by the following clauses for the \pftext{ltrans} predicate:
\begin{lstlisting}
Kind var_assoc type.
Type trans_to tm -> tm -> var_assoc.

Define ltrans : list var_assoc -> tm -> tm -> prop by
ltrans G X Y := exists G', select (trans_to X Y) G G' /\ no_elems G' ;
ltrans G (app M N) (app M' N') := exists G1 G2, 
    G ~ G1 ++ G2 /\ ltrans G1 M M' /\ ltrans G2 N N' ;
ltrans G (let T V E) (app (abs T E') V') := exists G1 G2, 
    G ~ G1 ++ G2 /\ ltrans G1 V V' /\ 
    nabla x y, ltrans (trans_to x y :: G2) (E x) (E' y) ;
ltrans G (abs T E) (abs T E') := 
    nabla x y, ltrans (trans_to x y :: G) (E x) (E' y).
\end{lstlisting}

We would like to prove that this translation preserves the types of
expressions.
Since translation and typing are defined by recursion over the
structures of expressions and will, in general, encounter open terms,
the theorem to be proved must have a form such as the
following: 
\begin{lstlisting}
Theorem ltrans_pres_ty'' : forall E E' T T' G G' G'',
mltype_of G E T -> ltrans G' E E' -> ltype_of G'' E' T' -> T = T'.
\end{lstlisting}
However, this formula cannot be proved as stated.
The contexts that arise at intermediate points in translation and type 
assignment have structures and relationships that must be made
explicit in the formulation to yield a provable statement.
Only names can be associated with other data in these contexts, and these
associations must be unique.
Further, we will need to relate the types of free variables in
a term and its translation to be able to show that the two have the
same type. 

The canonical way to make the relationship in the content of 
multiple contexts explicit in Abella is by defining an appropriate
context relation as a predicate.
Let \pftext{trans\_rel} be a predicate that encodes the relevant
relationship between the three contexts in consideration here.
The theorem to be actually proved then becomes the following:
\begin{lstlisting}
Theorem ltrans_pres_ty : forall E E' T T' G G' G'',
trans_rel G G' G'' -> mltype_of G E T -> ltrans G' E E' 
                   -> ltype_of G'' E' T' -> T = T'.
\end{lstlisting}
In proving theorems such as these, there are, once again, certain
lemmas about members of the contexts that we must be able to extract
from the relevant context relations.
In this particular example, we would need to be able to prove the
following lemmas that express a uniqueness property and a 
membership \emph{coordination} property between the related contexts:
\begin{lstlisting}
Theorem trans_rel_uniq : forall G1 G2 G3 X Y Y',
trans_rel G1 G2 G3 -> member (trans_to X Y) G2 
                   -> member (trans_to X Y') G2 -> Y = Y'.

Theorem trans_rel_mem : forall G1 G2 G3 E,
trans_rel G1 G2 G3 -> member E G2 -> exists X Y T, 
    E = trans_to X Y /\ name X /\ name Y /\ 
    member (ty_of X T) G1 /\ member (ty_of Y T) G3.
\end{lstlisting}
These properties are stated from the perspective of the second of the
three contexts.
There would be four more similar properties when matters are viewed
from either of the other two contexts.

The issue to be addressed, then, is how the context relation should be
defined to allow for the extraction of such properties.
There is a standard recipe for realizing the
described objectives when contexts are limited to a list-like structure.
In this example, we may define a list-oriented version of
\pftext{trans\_rel} following the conventional strategy as follows:
\begin{lstlisting}
Define trans_rel_list : list ty_assoc -> list var_assoc 
                                      -> list ty_assoc -> prop by
trans_rel_list nil nil nil ;
nabla x y, trans_rel_list (ty_of x T :: L1) 
                          (trans_to x y :: L2) 
                          (ty_of y T :: L3)   := trans_rel_list L1 L2 L3.
\end{lstlisting}
The uniqueness of binding property relativized to
\pftext{trans\_rel\_list} has a proof similar to the one
discussed for the typing context in Section~\ref{sec:bindingctx}.
The second property follows easily from the fact that the definition
is based on a coordinated recursion over the three contexts that in
fact ensures that they each contain the right kinds of members.

What we want, though, is a definition of \pftext{trans\_rel} that
applies to contexts whose structure includes the multiset union
constructor.
Using the ideas discussed in Section~\ref{sec:lifting}, we can
accomplish this once again by lifting the list-based definition up to 
contexts with a more general structure through permutations.
The following definition of the relation realizes the desired result:
\begin{lstlisting}
Define trans_rel : list ty_assoc -> list var_assoc -> list ty_assoc -> prop by
trans_rel G1 G2 G3 := exists L1 L2 L3, 
    G1 ~ L1 /\ G2 ~ L2 /\ G3 ~ L3 /\ trans_rel_list L1 L2 L3.
\end{lstlisting}
This definition still requires the associations in each context to
correspond with associations in the other contexts, but now the
corresponding associations need not be in the same position 
in each context.
Still, since the associations are clearly linked in the list-based
context relation, we will be able to lift the necessary membership
and uniqueness lemmas from the latter context relation.
Indeed, the proof of \pftext{trans\_rel\_mem} proceeds nearly as in the unary case: we
can make use of \pftext{mem\_replace} to ensure that \pftext{E} is an element of the
underlying translation context, and then make use of this lemma again to ensure that
\pftext{ty\_of X T} and \pftext{ty\_of Y T} are also members of the original typing
contexts.
The uniqueness lemma can be proved from the corresponding lemma for the
underlying context in a similar way, and the lifting process is even
simpler: since no conclusions need be drawn about the other contexts,
\pftext{mem\_replace} is only needed in one direction. 

The first clause in the definitions of \pftext{mltype\_of}, \pftext{ltrans},
and \pftext{ltype\_of} actually \emph{selects} an association from the
relevant context rather than simply checking membership.
Consequently, we would often need a stronger version of the
\pftext{trans\_rel\_mem} property that is based on the \pftext{select}
relation and that additionally asserts that the remaining contexts
continue to be in the \pftext{trans\_rel} relation:
\begin{lstlisting}
Theorem trans_rel_sel : forall G1 G2 G2' G3 E,
trans_rel G1 G2 G3 -> select E G2 G2' -> exists X Y T G1' G3', 
    E = trans_to X Y /\ name X /\ name Y /\ select (ty_of X T) G1 G1' /\
    select (ty_of Y T) G3 G3' /\ trans_rel G1' G2' G3'.
\end{lstlisting}
The new requirement here is that we must show that \pftext{trans\_rel
  G1' G2' G3'} holds for the three new contexts \pftext{G1'},
\pftext{G2'}, and \pftext{G3'} that result from selection from
\pftext{G1}, \pftext{G2}, and \pftext{G3}.
Most of this lemma can be proved without significant digression from
the proof sketched for \pftext{trans\_rel\_mem}.
For the lifting step, where we convert the \pftext{selects} on
multisets to \pftext{selects} on 
lists and vice versa, we can just use \pftext{sel\_replace} instead of
\pftext{mem\_replace}.
This also yields the necessary permutations for
concluding \pftext{trans\_rel G1' G2' G3'}: if \pftext{trans_rel G1 G2
  G3} holds because \pftext{trans_rel_list L1 L2 L3} does, and
selecting from \pftext{L1}, \pftext{L2}, and \pftext{L3} yields
\pftext{L1'}, \pftext{L2'}, and \pftext{L3'}, then \pftext{G2'~L2'},
\pftext{G1'~L1'},  and \pftext{G3'~L3'} must hold.
In the overall scheme, we can think of just proving \pftext{trans\_rel\_sel}.
We can get a proof of \pftext{trans\_rel\_mem} from this if it is
desired by using the following easily proved theorem that asserts that 
selecting from a context implies membership in that context:
\begin{lstlisting}
Theorem sel_implies_mem : forall X G G', select X G G' -> member X G. 
\end{lstlisting}

Finally, when multiset union is permitted in the construction of
contexts, we will need lemmas that verify the distributivity of
context relations over partitions.
For example, the definitions of \pftext{mltype\_of}, \pftext{ltrans},
and \pftext{ltype\_of} will force us to prove lemmas such as the
following that are analogous to the distributivity property for
\pftext{ty\_ctx\_distr} in the preceding section:\footnote{Note that
these lemmas do not let us specify the partition used for multiple
contexts at once; they assert only the existence of some partitions
that work. However, they suffice for many reasoning examples or can be
worked around by exploiting properties of other predicates---such as
the typing and translation relations here.}
\begin{lstlisting}
Theorem trans_rel_distr : forall G1 G1' G1'' G2 G3,
trans_rel G1 G2 G3 -> G1 ~ G1' ++ G1'' -> exists G2' G2'' G3' G3'',
    G2 ~ G2' ++ G2'' /\ G3 ~ G3' ++ G3'' /\
    trans_rel G1' G2' G3' /\ trans_rel G1'' G2'' G3''.
\end{lstlisting}

To prove such a distributivity lemma, we can first state and prove an
analogous lemma for the related contexts in list form and then lift it to
contexts with a more general structure.
For this, we may reuse the definition of \pftext{partition}
and many of its properties, stating the lemma to prove in the case
under consideration as 
\begin{lstlisting}
Theorem trans_rel_list_distr : forall L1 L1' L1'' L2 L3,
trans_rel_list L1 L2 L3 -> partition L1 L1' L1'' -> exists L2' L2'' L3' L3'',
    trans_rel_list L1' L2' L3' /\ trans_rel_list L1'' L2'' L3'' /\
    partition L2 L2' L2'' /\ partition L3 L3' L3''.
\end{lstlisting}
The ordered nature of \pftext{partition} again is critical to the proof; since the
related contexts are also ordered, we can construct new \pftext{partitions} using the
corresponding elements as in \pftext{partition L1 L1' L1''} in the
same places for the other contexts.
Then, to lift this lemma to arbitrary multiset partitions, we can
exploit \pftext{perm\_to\_part} in a first step to transform
\pftext{G1~G1'++G1''} into  \pftext{partition L1 L1' L1''}, where
\pftext{G1~L1}, \pftext{G1'~L1'}, and \pftext{G1''~L1''} hold.
After applying \pftext{trans\_rel\_list\_distr}, we can make use 
of a kind of inverse of the \pftext{perm\_to\_part} lemma to convert
\pftext{partitions} back into permutations:
\begin{lstlisting}
Theorem part_to_perm : forall L L1 L2, partition L L1 L2 -> L ~ L1 ++ L2.
\end{lstlisting}
Since partitioning a list involves a restricted form of selection, the
structure of the proof of this lemma should be easy to visualize.
To complete the proof of \pftext{trans\_rel\_distr}, we can then note that the 
contexts it asserts the existence of can be the same as those asserted
by \pftext{trans\_rel\_list\_distr}, and that for any \pftext{G},
\pftext{L}, \pftext{L'}, and \pftext{L''},  \pftext{G~L'++L''} follows from
\pftext{G~L} and \pftext{L~L'++L''} by properties of \pftext{perm}. 
Thus, we can conclude that \pftext{trans\_rel G1' G2' G3'} and
\pftext{trans\_rel G1'' G2'' G3''} hold by definition; we will need to
show that each list is a permutation of itself for this, but  this
follows easily from the fact that each is a list. 

\section{Schematic Context Specifications and Automated Proofs}\label{sec:auto}

The idea of defining a multiset-based context
specification---via a context predicate or context relation---by
lifting from a list-based one has a general applicability and can be
deployed in other developments as well.
We present in this section a general form for such specifications for
which we can write \emph{schematic} proofs of several distributivity
lemmas and of a lifting procedure for a reasonably large class of
lemmas based on the \pftext{member} predicate which includes our
membership and uniqueness lemmas.
This works since the distributivity lemmas and lifting
procedure depend only on the general structure of the context
specifications defined and not on the particular elements of the
context(s).
Hence, a user need only state and prove the
\pftext{member} lemmas that require explicit reference to the elements
of the binding context(s) for an underlying specification and can
leave the rest of the work to an automated procedure.

Let us begin by introducing a command that might be used
to succinctly generate a pair of context specifications---one based on
lists and the other based on multisets. The syntax of this command
should take the following general form, with each \pftext{FORMULA}
referring to an expression of type \pftext{prop}, each
\pftext{TERM} referring to a term of some other type, each
\pftext{VAR} referring to a variable identifier, and \pftext{CTX-NAME}
referring to the name of the context specification to be defined:
\begin{lstlisting}
Context CTX-NAME with elems as
    nabla VAR11 ... VAR1k_1 (TERM11 _|_ ... _|_ TERM1n -| FORMULA1) \/ ... \/
    nabla VARm1 ... VARmk_m (TERMm1 _|_ ... _|_ TERMmn -| FORMULAm).
\end{lstlisting}
The use of the formula is to provide an additional means for encoding
the relationship between elements of each of the related contexts in a
context relation.
Also note that there may be zero variables, in which case the
\pftext{nabla} may be omitted, and we can omit the formula if it is
\pftext{true}.
However, there must be at least one clause and at least one
term denoting some element of a context.
As examples of the intended
usage of this command, we present the commands that, following the
process we describe next, would generate the
definitions of \pftext{ty\_ctx'} and \pftext{trans\_rel} from
Sections~\ref{sec:lifting} and \ref{sec:ctxrels}:
\begin{lstlisting}
Context ty_ctx' with elems as nabla x (ty_of x T).
Context trans_rel with elems as
    nabla x y (ty_of x T _|_ trans_to x y _|_ ty_of y T').
\end{lstlisting}

Our command schema is meant to define a pair of context specifications of the
following forms, where each \pftext{TYPE} is an Abella type inferred from the types
of each contexts' elements:
\begin{lstlisting}
Define CTX-NAME_list : list TYPE1 -> ... -> list TYPEn -> prop by
CTX-NAME_list nil ... nil ;
nabla VAR11 ... VAR1k_1, CTX-NAME_list (TERM11 :: L1) ... (TERM1n :: Ln) :=
    CTX-NAME_list L1 ... Ln /\ FORMULA1 ;
...
nabla VARm1 ... VARmk_m, CTX-NAME_list (TERMm1 :: L1) ... (TERMmn :: Ln) := 
    CTX-NAME_list L1 ... Ln /\ FORMULAm.

Define CTX-NAME : list TYPE1 -> ... -> list TYPEn -> prop by
CTX-NAME G1 ... Gn := exists L1 ... Ln,
    G1 ~ L1 /\ ... /\ Gn ~ Ln /\ CTX-NAME_list L1 ... Ln.
\end{lstlisting}

Once we have a context specification of the aforementioned form, a
suite of lemmas can be automatically generated about it.
First, a distributivity lemma can be generated for each index of the
specification that allows the context specification to be
distributed over partitions of the corresponding context while
generating corresponding partitions of the other context(s) as needed
for the other indices.
The general form of the ith such lemma may be represented as follows:
\begin{lstlisting}
Theorem CTX-NAME_distri : forall G1 ... Gi Gi' Gi'' ... Gn,
CTX-NAME G1 ... Gn -> Gi ~ Gi' ++ Gi'' -> exists G1' G1'' ... Gn' Gn'',
    CTX-NAME G1' ... Gn' /\ CTX-NAME G1'' ... Gn'' /\ 
    G1 ~ G1' ++ G1'' /\ ... /\ Gn ~ Gn' ++ Gn''.
\end{lstlisting}
For instance, for \pftext{trans\_rel}, we might automatically generate the lemma for i = 2 in
this form as:
\begin{lstlisting}
Theorem trans_rel_distr2 : forall G1 G2 G2' G2'' G3,
trans_rel G1 G2 G3 -> G2 ~ G2' ++ G2'' -> exists G1' G1'' G3' G3'',
    trans_rel G1' G2' G3' /\ trans_rel G1'' G2'' G3'' /\  
    G1 ~ G1' ++ G1'' /\ G3 ~ G3' ++ G3''.
\end{lstlisting}
Each lemma can be proved automatically as well. The generated proofs follow
the structure that we have already seen in Sections~\ref{sec:lifting} and \ref{sec:ctxrels}.
In short, a corresponding lemma involving \pftext{partition} is first automatically
generated and proved by a routine inductive argument that depends only on the number
of clauses and names in the definition. Then, \pftext{perm\_to\_part} is applied to
interface the desired lemma's hypotheses with this lemma's, and finally 
\pftext{part\_to\_perm} is applied to obtain results in the right form.

For lemmas involving \pftext{member}, an algorithm exists to automatically lift lemmas
proved for traditional context specifications to their multiset versions. Suppose
the user proves a lemma of the following form:
\begin{lstlisting}
Theorem USER-LEMMA : forall L1 ... Ln VAR*,
CTX-NAME_list L1 ... Ln -> [member TERM Li ->]* [exists VAR*, ] 
    [member TERM Lj /\]* [FORMULA /\]* [TERM = TERM /\]* true.
\end{lstlisting}
Suppose also that each \pftext{FORMULA} and \pftext{TERM} does not depend on any of 
the context variables \pftext{Li}, so that any non-\pftext{member} assertions are only
about the elements of the context(s). Then, a corresponding lemma for multiset-based 
contexts, of the following form, can be automatically generated and proved:
\begin{lstlisting}
Theorem USER-LEMMA-MSET : forall G1 ... Gn VAR*,
CTX-NAME G1 ... Gn -> [member TERM Gi ->]* [exists VAR*, ] 
    [member TERM Gj /\]* [FORMULA /\]* [TERM = TERM /\]* true.
\end{lstlisting}
The automatically generated proof involves three main steps:
\begin{enumerate}
\item The context specification hypothesis \pftext{CTX-NAME G1 ... Gn}
  is unfolded and appropriate instances of \pftext{mem\_replace} are
  applied to each of the other hypotheses.

\item The user-provided lemma is applied to the hypotheses constructed
  in the first step.
  
\item The conclusions obtained using the user-provided lemma are
  converted into the desired forms. Nothing needs to be done for the
  \pftext{FORMULA} and equality conclusions, but any obtained
  instances of \pftext{member} are converted to the correct form via
  appropriate uses of \pftext{mem\_replace}. Since the original
  lemma's form was restricted to only allow \pftext{member} to pull
  from contexts described by the context specification, it is certain
  that the requisite permutations will be available; they are
  necessarily the same permutations as obtained by unfolding
  \pftext{CTX-NAME G1 ... Gn}.
\end{enumerate}

We envision tactics in Abella for handling the aforementioned
automation.  A \pftext{subst} tactic would implement the
\pftext{mem\_replace} lemma, a 
\pftext{distr} tactic would implement the distributivity lemmas, and a
\pftext{lift} tactic would implement the
procedure for lifting \pftext{member}-based lemmas. 
For example, calling \pftext{subst Hi into Hj}
would apply the \pftext{mem\_replace} lemma, as long as \pftext{Hi} is
an appropriate permutation and \pftext{Hj} is an instance of the
\pftext{member} predicate. On the other hand, calling \pftext{distr Hi
  over Hj} with \pftext{Hi} being a context specification defined via
the \pftext{Context} command and \pftext{Hj} being a
\pftext{perm}-style partition would prove and apply a distributivity
lemma for whichever index of the context specification could be
matched with the input permutation.  And, finally, calling
\pftext{lift USER-LEMMA} would generate and prove the corresponding
\pftext{USER-LEMMA-MSET}, adding it as a new hypothesis.

\section{Related Work}\label{sec:related}

Our focus in this paper has been on the special problems that arise
when binding contexts must be accorded a resource interpretation.
While this concern is original to our work, we have superimposed it on
a treatment of resources, which is an issue that has received the
attention of other researchers.
A particular situation in which the need for such a treatment has
arisen is in the encoding of linear logic~\cite{girard87tcs}
within proof assistants towards mechanizing reasoning about the
meta-theoretic properties of this logic.
Chaudhuri \etal\ have undertaken this task using the Abella
system~\cite{chaudhuri19tcs}.  
They too have used multisets to encode resources, which, in their
case, are linear collections of formulas.
They observe that the representation of multisets must support
the ability to add an element to a multiset and to partition a
multiset, and it must ensure that multisets are considered to be
equivalent under permutations. 
These observations underlie our work as well, with the key difference
that we have taken permutation equivalence to be fundamental to the
representation. 
This allows us to introduce the \munion{} constructor that renders
partitioning into a syntactic operation rather than needing it to be
defined, as is done by the \pftext{merge} predicate in~\cite{chaudhuri19tcs}.
Our approach has the benefit of succinctness, at least in
presentation; for example, it accommodates a simple rendition of the
partitioning of a multiset into several subcomponents.
On the negative side, the definitions of permutation and the addition (or,
dually, the selection) of an element are marginally more complex.
Similar concerns arise in the encoding of linear logic in the Coq
system developed by Olivier Laurent~\cite{laurent18yalla}.
In that work, the choice was made to use lists to represent linear
collections of formulas and to realize the multiset interpretation via
an explicit ``exchange'' rule that is implemented via permutations.
While this approach supports a simple encoding, it separates
partitioning from permutations, an aspect that can make the analysis
of the derivability of particular sequents in linear logic more
complex.

The notion of partitionable contexts is also relevant to 
the {\sc Lincx} framework~\cite{georges17esop} that allows the user to define 
functions whose types correspond to typing judgments in the linear
logical framework LLF~\cite{cervesato02ic}; theorems given by the type
of the function are considered proved if the function can be shown to
be total.
Unlike our scheme that uses the explicit definition of a permutation
relation for treating partitions, {\sc Lincx} provides a built-in
operator $\bowtie$ for \emph{context joins}, whose definition is
hidden from the user.
One significant difference between these schemes is that, since LLF contexts
are inherently ordered, context joins must preserve the relative order of
elements whereas \pftext{perm}-style partitions need not.
Each element of \pftext{G = G1} $\bowtie$ \pftext{G2} remains in the same order in
\pftext{G1} and \pftext{G2} but is only made \emph{available} in exactly one 
of them---with only a placeholder in the other for order-preservation and type 
checking purposes.
Since context joins are built-in and system-manipulated, a user need
not explicitly drive  the functionality of these.
However, by the same token, they also cannot affect the functionality.
In contrast, we expose the definition of \pftext{perm} and allow users
to reason about it and prove additional lemmas if needed---though the user
also typically \emph{must} reason explicitly about \pftext{perm} in order
to make use of it.

The encoding that we have used for type assignment in the simply typed
linear lambda calculus is based on superimposing linearity explicitly on
typing contexts. 
This choice has been motivated by the eventual application for our
work that we discuss in the next section; the simply typed linear
lambda calculus figures mainly as an example to highlight the issues
that have to be considered in this setting.
If the focus is instead on a specific example,
then an encoding of a different style could be used to circumvent the
issues discussed.
For instance, the simply typed linear lambda calculus could have been
treated by specifying type assignment and linearity separately;
uniqueness of typing in this case would, for example, be a simple
consequence of the result for the regular simply typed lambda
calculus. 
This style in fact underlies the encoding of linear logic and other
substructural logics described in~\cite{crary10icfp}. 

The idea of schematically extracting context properties, useful for minimizing
the burden of reasoning explicitly about contexts, has also been explored in 
other settings besides ours. Savary B\'elanger and Chaudhuri~\cite{belanger14lfmtp}
define a plugin for Abella for concisely defining and extracting properties from what
they call \emph{regular context relations}, which describe the structure of 
LF contexts. This structure is noticeably similar to the structure of context
specifications without the addition of the lifting procedure using \pftext{perm}. 
Specifically, in their framework, a user can define a \emph{context schema} 
that fully specifies the form of the elements in the desired context(s). Then, 
they can make use of provided \emph{tacticals} for extracting properties of 
the corresponding context relation. For example, the \emph{inversion} tactical 
extracts the form of (corresponding) elements in the context(s), much like our 
membership lemmas. Though both our and their developments define a
general form for contexts of interest that capture some desired
properties and then provide tools for extracting those properties, the
specific goals and the actual form of the contexts differ significantly.

\section{Conclusion}\label{sec:concl}

In this paper, we have discussed our scheme for specifying binding
contexts that must be partitionable in Abella.
We have illustrated our ideas by using typing judgments and a
translation relation that are encoded directly as definitions in the
\emph{reasoning logic} of Abella. 
However, reasoning in Abella is often done using a two-level logic
approach, in which the reasoning logic is augmented with an
auxiliary \emph{specification logic} that is well-suited for computation.
This paradigm is realized by embedding the specification logic in the
reasoning logic via a definition that encapsulates the proof system of
the specification logic; one then describes object systems in the
specification logic and reasons about them via the ability the
embedding provides to reason about derivability in the specification
logic.
In ongoing work, we are exploring the possibility of using a variant
of linear logic called
\emph{Forum}~\cite{miller96tcs} that has a computational
interpretation as the specification logic in this framework.
The motivation for doing so is that linear object systems, such as the
linear lambda calculus that we have considered here, can be specified
in a logical way by making use of \emph{linear implication}
($\multimap$) to encode resources and their usage, a move that enables
metatheoretic properties of the specification logic to be used in
simplifying the reasoning process. 
To support this idea, we must provide an embedding of Forum in Abella.
Since formulas in one category in such a logic must
be used exactly once, their encoding and usage in the embedding 
necessitates a treatment of linear contexts with an associated
capability for considering their partitioning in the reasoning
process.\footnote{Note that the kind of embedding we are interested in
here makes it necessary to treat linear contexts explicitly, unlike
what is done, for example, in \cite{crary10icfp}.}
Moreover, when the object system embodies notions of binding, such
linear contexts take on the attributes of binding contexts that have
been the topic of interest in this paper. 
Many of the ideas we have discussed remain applicable in this
situation and we are in fact incorporating the automation techniques
described in Section~\ref{sec:auto} in our implementation towards
providing the user a tool to simplify reasoning developments that 
make use of the new specification logic. 

\section*{Acknowledgements}

This paper has benefitted greatly from the feedback provided by its
reviewers.
The work underlying it was supported in its early stages by the
National Science Foundation under Grant No. CCF-1617771.
Any opinions, findings, and conclusions or recommendations expressed
in this material are those of the authors and do not necessarily
reflect the views of the National Science Foundation.

\bibliographystyle{eptcs}
\bibliography{lfmtp}{}

\end{document}